\documentstyle[12pt,graphicx,pstricks,braket,amsmath,amssymb,enumitem,
                    epstopdf,pdfpages,cite,tabu,hyperref]{article}
\numberwithin{equation}{section}
\begin{document}

\def\AEF{A.E. Faraggi}

\def\JHEP#1#2#3{{JHEP} {\textbf #1}, (#2) #3}
\def\vol#1#2#3{{\bf {#1}} ({#2}) {#3}}
\def\NPB#1#2#3{{\it Nucl.\ Phys.}\/ {\bf B#1} (#2) #3}
\def\PLB#1#2#3{{\it Phys.\ Lett.}\/ {\bf B#1} (#2) #3}
\def\PRD#1#2#3{{\it Phys.\ Rev.}\/ {\bf D#1} (#2) #3}
\def\PRL#1#2#3{{\it Phys.\ Rev.\ Lett.}\/ {\bf #1} (#2) #3}
\def\PRT#1#2#3{{\it Phys.\ Rep.}\/ {\bf#1} (#2) #3}
\def\MODA#1#2#3{{\it Mod.\ Phys.\ Lett.}\/ {\bf A#1} (#2) #3}
\def\RMP#1#2#3{{\it Rev.\ Mod.\ Phys.}\/ {\bf #1} (#2) #3}
\def\IJMP#1#2#3{{\it Int.\ J.\ Mod.\ Phys.}\/ {\bf A#1} (#2) #3}
\def\nuvc#1#2#3{{\it Nuovo Cimento}\/ {\bf #1A} (#2) #3}
\def\RPP#1#2#3{{\it Rept.\ Prog.\ Phys.}\/ {\bf #1} (#2) #3}
\def\APJ#1#2#3{{\it Astrophys.\ J.}\/ {\bf #1} (#2) #3}
\def\APP#1#2#3{{\it Astropart.\ Phys.}\/ {\bf #1} (#2) #3}
\def\EJP#1#2#3{{\it Eur.\ Phys.\ Jour.}\/ {\bf C#1} (#2) #3}
\def\etal{{\it et al\/}}
\def\notE6{{$SO(10)\times U(1)_{\zeta}\not\subset E_6$}}
\def\E6{{$SO(10)\times U(1)_{\zeta}\subset E_6$}}
\def\highgg{{$SU(3)_C\times SU(2)_L \times SU(2)_R \times U(1)_C \times U(1)_{\zeta}$}}
\def\highSO10{{$SU(3)_C\times SU(2)_L \times SU(2)_R \times U(1)_C$}}
\def\lowgg{{$SU(3)_C\times SU(2)_L \times U(1)_Y \times U(1)_{Z^\prime}$}}
\def\SMgg{{$SU(3)_C\times SU(2)_L \times U(1)_Y$}}
\def\Uzprime{{$U(1)_{Z^\prime}$}}
\def\Uzeta{{$U(1)_{\zeta}$}}

\newcommand{\cc}[2]{c{#1\atopwithdelims[]#2}}
\newcommand{\bev}{\begin{verbatim}}
\newcommand{\beq}{\begin{equation}}
\newcommand{\ba}{\begin{eqnarray}}
\newcommand{\ea}{\end{eqnarray}}

\newcommand{\beqa}{\begin{eqnarray}}
\newcommand{\beqn}{\begin{eqnarray}}
\newcommand{\eeqn}{\end{eqnarray}}
\newcommand{\eeqa}{\end{eqnarray}}
\newcommand{\eeq}{\end{equation}}
\newcommand{\beqt}{\begin{equation*}}
\newcommand{\eeqt}{\end{equation*}}
\newcommand{\Eev}{\end{verbatim}}
\newcommand{\bec}{\begin{center}}
\newcommand{\eec}{\end{center}}
\newcommand{\bes}{\begin{split}}
\newcommand{\ees}{\end{split}}
\def\ie{{\it i.e.~}}
\def\eg{{\it e.g.~}}
\def\half{{\textstyle{1\over 2}}}
\def\nicefrac#1#2{\hbox{${#1\over #2}$}}
\def\third{{\textstyle {1\over3}}}
\def\quarter{{\textstyle {1\over4}}}
\def\m{{\tt -}}
\def\mass{M_{l^+ l^-}}
\def\p{{\tt +}}

\def\slash#1{#1\hskip-6pt/\hskip6pt}
\def\slk{\slash{k}}
\def\GeV{\,{\rm GeV}}
\def\TeV{\,{\rm TeV}}
\def\y{\,{\rm y}}

\def\l{\langle}
\def\r{\rangle}
\def\LRS{LRS  }

\begin{titlepage}
\samepage{
\setcounter{page}{1}
\rightline{LTH--1052} 
\rightline{MAN/HEP/2015/13}
\vspace{1.5cm}

\begin{center}
 {\Large \bf 
Extra $Z^\prime$s and $W^\prime$s in\\ \medskip 
Heterotic--String Derived Models}
\end{center}

\begin{center}
%\vspace{1.cm}

Alon E. Faraggi$^{\dagger}$\footnote{
		                  E-mail address: alon.faraggi@liv.ac.uk}
and 
Marco Guzzi$^{\ddagger}$\footnote{ E-mail address: marco.guzzi@manchester.ac.uk}
\\
\vspace{.25cm}
{\it $^{\dagger}$Department of Mathematical Sciences\\
University of Liverpool, Liverpool, L69 7ZL, United Kingdom}\\
\vspace{.05in}
{\it $^{\ddagger}$Consortium for Fundamental Physics,\\
School of Physics \& Astronomy\\
University of Manchester, Manchester, M13 9PL, United Kingdom}
\end{center}

\begin{abstract}

The ATLAS and CMS collaborations recently recorded possible excess in 
the di--boson production at the di--boson invariant mass at 
around 2 TeV. Such an excess may be produced if there exist
additional $Z^\prime$ and/or $W^\prime$ at that scale. 
We survey the
extra $Z^\prime$s and $W^\prime$s that may
arise from semi--realistic heterotic string vacua
in the free fermionic formulation in seven distinct cases
including: 
$U(1)_{Z^\prime}\in SO(10)$;  
family universal $U(1)_{Z^\prime}\notin SO(10)$;  
non--universal $U(1)_{Z^\prime}$;
hidden sector $U(1)$ symmetries and kinetic mixing;
left--right symmetric models;
Pati--Salam models; 
leptophobic and custodial symmetries. 
Each case has a distinct signature 
associated with the extra symmetry breaking scale.
In one of the cases we explore the discovery potential 
at the LHC using resonant leptoproduction. 
Existence of extra vector boson with the reported properties 
will significantly constrain the space of allowed string vacua.

\end{abstract}
\smallskip}
\end{titlepage}

\section{Introduction}

The Standard Model multiplet structure strongly favours its 
embedding in chiral 16 representations of $SO(10)$.
This can be most emphatically demonstrated by
remembering that the Standard Model gauge charges
are experimental observables. The Standard Model, 
including right--handed neutrinos, 
has three group factors, three generations, and six 
multiplets per family, and therefore heuristically
the number of parameters required in the Standard Model
is fifty--four. Embedding the Standard Model states 
in $SO(10)$ representations reduces this number to one, 
which is the number of 16 spinorial $SO(10)$ representations
required to accommodate the Standard Model states. 

Gravitational interactions are not accounted for in
the Standard Model. A contemporary self--consistent
framework that facilitates the exploration of the 
synthesis of the gravitational and gauge interactions 
is provided by string theories, which are conjectured
to be effective limits of
a more fundamental theory. Hetertic--string theory
is the perturbative limit that allows for the embedding
of the Standard Model states in chiral $SO(10)$ 
representations as it gives rise the spinorial $16$
representations in its perturbative spectrum. Three
generation models with viable gauge group and Higgs
states have been constructed using a variety of methods.
Among those the free fermionic formulation \cite{fff} of the
heterotic string \cite{heterotic}
provided a particularly fertile ground.
In these three generation models the $SO(10)$ symmetry 
is broken at the string level to one of its maximal 
subgroups. 

Recently, the ATLAS and CMS collaborations \cite{atlas,cms-1,cms-2} reported an excess
in fat jet production which is kinematically compatible
with the decay of a heavy resonance into two vector bosons, 
generating a wide range of interest \cite{dl}.
A possible interpretation of the observed excess is as 
an extra $Z^\prime$ or $W^\prime$ with a mass of the order 
of few TeV's \cite{atlas,cms-1,dl}. 
The existence of an extra $Z'$ inspired 
from heterotic string theory attracted considerable interest
in the particle physics literature \cite{e6zprime}. However, 
constructing string models that allow an $Z'$ to remain
unbroken down to low scales has proven to be very challenging. 
The reason being that the extra $U(1)$ symmetries that are
studied in the literature are either anomalous or 
have to be broken at the high scale
to generate qualitatively realistic fermion mass spectrum.
Furthermore, Flavour Changing Neutral Current (FCNC) 
Constraints, indicate that the extra $Z^\prime$ below the DecaTeV 
scale, has to be family universal, and imposes an 
additional strong constraint on the viable string vacuum. 
Extra vector bosons in the TeV region will exclude the 
majority of heterotic--string models constructed to date. 
Recently, a semi--realistic string derived model that allow
for a light $Z^\prime$ model was constructed in ref. \cite{fr}.

In this paper we survey the various types of extra $Z^\prime$s
that may arise from heterotic string models. Our 
laboratory to examine this question is provided
by the three generation heterotic string models 
in the free fermionic formulation. These
class of string vacua is related to $Z_2\times Z_2$ 
orbifold compactification \cite{z2xz2}, but the properties of the 
models pertaining to the gauge group structure are
relevant to other constructions \cite{others}. 
The various $Z^\prime$ that arise in the models
may be classified into several broad categories:
\begin{itemize}
\item
Family universal $U(1)$s that admit the 
$SO(10)$ and $E_6$ embedding of the Standard Model 
charges.
\item Extra $W^\prime$ \& $Z^\prime$ arising in left--right symmetric 
heterotic--string models. 
\item Family non--universal $U(1)$s.
\item Hidden sector $U(1)$ symmetries and kinetic mixing. 
\item Extra vector bosons from extensions of the colour group. 
\item Leptophobic and custodial $SU(2)$ symmetries.
\end{itemize}
We will comment on these possibilities and their viability below the 
10 TeV scale. We show that extra hidden sector $U(1)$s is not a viable
possibility in these models. Furthermore, each of the remaining 
cases carries a unique signature associated with extra gauge 
symmetry breaking scale. For example, $U(1)_{Z^\prime}\in SO(10)$ 
only requires additional right--handed neutrinos for anomaly cancellation, 
whereas $U(1)_{Z^\prime}\notin SO(10)$ mandates the existence of additional
matter. Family non--universal $U(1)$s are constrained to be above
the DecaTeV scale, whereas non--Abelian extensions of the
Standard Model gauge symmetries, as in the left--right symmetric
models, give rise to additional vector bosons. Discovery 
of one or more additional vector bosons at the LHC will
therefore pave the way to discriminate between the different
possibilities and will strengthen the case for a multi--TeV 
lepton collider and a 100 TeV hadron collider. In section 
\ref{lhcprop} we explore the discovery potential 
at the LHC using resonant leptoproduction. 

\section{Additional $U(1)$s in heterotic--string
models}\label{au1}

In this section we elaborate on the type 
of extra gauge bosons that may arise from heterotic--string 
vacua. Our discussion is in the framework of the free fermionic formulation. 
Details of the construction and the models that we discuss are given in the 
the references provided, and will not be repeated here.
In this paper we only mention the features that are relevant 
for the discussion of the light extra $W^\prime$s and
$Z^\prime$s, which are obtained from the untwisted 
Neveu--Schwarz sector. The last category that we consider includes
vector bosons from additional sectors. 

In the free fermionic formulation of the heterotic--string all the 
degrees of freedom needed to cancel the conformal 
anomaly are represented in terms of world--sheet fermions 
propagating on the string world--sheet. In the light--cone gauge 
in four dimensions 20 right--moving and 44 left--moving world--sheet
real fermions are required. These are typically denoted as 
$$
\{\psi^{1,2},\chi^{1, \cdots  ,6}, y^{1, \cdots  ,6}, \omega^{1, \cdots  ,6}
~\vert~
 {\bar y}^{1, \cdots  ,6}, {\bar\omega}^{1, \cdots  ,6},
{\bar\psi}^{1, \cdots ,5} ,{\bar\eta}^{1,2,3},{\bar\phi}^{1,\cdots,8}\}
$$
where in the right--moving bosonic sector
the $\{{\bar y}^{1, \cdots  6}, {\bar\omega}^{1, \cdots  6}\}$ 
are real and 
$\{{\bar\psi}^{1, \cdots ,5}, {\bar\eta}^{1,2,3},{\bar\phi}^{1,\cdots,8}\}$
are complex. A complex world--sheet fermion produces a
$U(1)$ current in the Cartan subalgebra of the string models. 
The 16 complex right--moving world sheet fermions therefore generate
a rank 16 gauge group. Additional Cartan generators in the 
four dimensional gauge group may be obtained by complexifying
additional world--sheet fermions from the set 
$\{{\bar y}^{1, \cdots  6}, {\bar\omega}^{1, \cdots  6}\}$. 
The five world--sheet complex fermions 
${\bar\psi}^{1,\cdots,5}$ are the Cartan generators of 
the $SO(10)$ gauge symmetry and ${\bar\eta}^{1,2,3}$ 
generate three $U(1)$ symmetries in the observable sector, 
denoted by $U(1)_{1,2,3}$. 
The three generation
free fermionic models typically contain 
up to three additional $U(1)$ symmetries from the 
set of real fermions $\{{\bar y}^{1, \cdots  6}, {\bar\omega}^{1, \cdots  6}\}$,
denoted by $U(1)_{4,5,6}$. The symmetries discussed up to know are 
all in the observable sector, whereas the eight 
complex world--sheet fermions ${\bar\phi}^{1,\cdots,8}$
correspond to the Cartan generators of the hidden sector
gauge group. The distinction between hidden and observable 
entails that the states that are identified as the 
Standard Model states may carry charges under the
observable gauge symmetries but may not carry 
hidden charges. 

Under parallel transport around the noncontractible 
loops of the world--sheet torus
of the vacuum to vacuum amplitude, the world--sheet fermions 
pick up a phase. The allowed phase assignments are constrained
by the requirement that the vacuum to vacuum amplitude is invariant 
under modular transformations. Models in the free fermionic formulation
are obtained by specifying a set of boundary basis vectors and 
the associated one--loop GGSO phases \cite{fff}, 
which both must satisfy a set of constraints derived 
by the requirement that the vacuum to vacuum amplitude is
invariant under modular transformations. In this paper 
we will focus on the so--called NAHE--based models \cite{nahe}, 
which are typically produced by a set of eight (or nine)
boundary condition basis vectors denoted by 
$\{{\bf 1}, S, b_1, b_2, b_3, \alpha, \beta, \gamma\}$,
where the set $\{{\bf 1}, S, b_1, b_2, b_3\}$ is the 
so--called NAHE--set \cite{nahe}. The basis vectors of the NAHE--set
preserve the $SO(10)$ symmetry. Basis vectors that extend
the NAHE--set may preserve the $SO(10)$ symmetry
in which case they are denoted as $b_{4, 5, \cdots}$,
or they may break the $SO(10)$ symmetry, in 
which case they are denoted as $\{\alpha, \beta, \gamma, \cdots\}$. 
At least one basis vector beyond the NAHE--set must 
break the $SO(10)$ symmetry. 

Space--time vector bosons in the free fermionic models arise
from the untwisted Neveu--Schwarz sector and possibly from 
additional sectors that are obtained from combinations 
of the basis vectors. The vector bosons from these
additional sectors enhance the gauge symmetry which
is obtained from the untwisted NS--sector. 
The generators of the $SO(10)$ symmetry and of any additional
$U(1)$ symmetries are obtained from the untwisted 
NS--sector. The vector bosons arising in the additional
sectors do not play a role in the case of extra gauge symmetries
from $SO(10)$ subgroups, or from extra NS $U(1)$ symmetries. 
They arise in the case of custodial symmetries \cite{custsu2}.
The projection of the spacetime vector bosons arising 
from the untwisted NS sector depend only on the boundary condition 
basis vectors, and do not depend on the GGSO phases \cite{fff}.
The type of enhancement from the additional sectors does depend 
on the GGSO phases, but it will not play a role in our discussion here. 
The boundary condition basis 
vectors, and the
GGSO phases, leading to the models that we discuss
are given in the references. 

The three sectors $b_{1,2,3}$ correspond to the three twisted
sectors of the $Z_2\times Z_2$ orbifold. The basis vector
$S$ is the spacetime supersymmetry generator, and insures the 
projection  of the untwisted NS tachyon. 
At the level of the NAHE--set each of the twisted sectors
produces sixteen multiplets in the $16$ spinorial representation
of $SO(10)$. The additional basis vectors beyond the NAHE--set
reduce the number of generations to three generations and at the 
same time break the $SO(10)$ symmetry to one of its maximal subgroups. 
Semi--realistic models were obtained with: 
\begin{itemize}
\item $SU(5)\times U(1)$ (FSU5) \cite{fsu5};
\item $SU(3)\times SU(2)\times U(1)^2$ (SLM) \cite{fny,slm};
\item $SO(6)\times SO(4)$ (PS) \cite{alr}
\item $SU(3)\times U(1) \times SU(2)^2$ (LRS) \cite{lrs},
\end{itemize}
whereas the $SU(4)\times SU(2)\times U(1)$ (SU421) 
class of models has been shown not to produce phenomenologically
realistic examples \cite{su421}. 

All of the three generation free fermionic models share a common 
structure due to the underlying $SO(10)$ symmetry and the 
spectrum available to break the $U(1)$ symmetry
which is embedded in $SO(10)$ and is orthogonal to the
weak hypercharge. 
In all these models this extra $U(1)$ symmetry
is necessarily broken by a Higgs field with
charges identical to those of the right--handed neutrino, 
{\it i.e.} the Standard Model singlet that resides 
in the $16$ spinorial representation of $SO(10)$.
The reason is the absence of the adjoint
and higher level representations in the massless spectrum of
these string models. 
All the semi--realistic models contain 
three chiral 16 representations of $SO(10)$ decomposed 
under the final $SO(10)$ subgroup and electroweak Higgs 
doublet representations that arise from the vectorial 
10 representation of $SO(10)$. 

One distinction between the models is the scale
at which the $SO(10)$ extra $U(1)$ has to be broken. 
For instance in the case of the FSU5 models it
must be broken at the MSSM GUT scale, to generate 
masses for the $SU(5)\times U(1)$ vector bosons 
which mediate proton decay via dimension six operators. 
In the three other cases it could in principle remain 
unbroken below that scale, because these models do not contain 
vector bosons that may mediate proton decay 
via dimension six operators. 

Another distinction between the models is with respect to 
the anomalous $U(1)$ symmetry that arises in the string models
\cite{cleaverau1}. 
In the case of the FSU5, SLM and PS models, 
the $U(1)_{1,2,3}$ symmetries, as well as their linear combination, 
\beq
U(1)_\zeta= U(1)_1+U(1)_2+U(1)_3 
\label{u1zeta}
\eeq
are anomalous, whereas in the LRS and SU421 models 
they are anomaly free. In the models in which this
$U(1)$ symmetry is anomalous it is broken by the 
Dine--Seiberg--Witten anomaly 
cancellation mechanism \cite{DSW}, 
whereas in models in which it is anomaly free it could 
in principle remain unbroken down to low scales. 
The basic characteristic of the FSU5, SLM and PS 
cases in this regard is that they emanate from the 
symmetry breaking pattern $E_6\rightarrow SO(10)\times U(1)_\zeta$,
induced by the GGSO projections. In this case 
$U(1)_\zeta$ becomes anomalous because the $10+1 $
components in the 27 representation of $E_6$ 
are projected out, resulting in $U(1)_\zeta$ becoming
anomalous. The LRS \cite{lrs} and SU421 models \cite{su421}
circumvent the $E_6\rightarrow SO(10)\times U(1)_\zeta $
symmetry breaking pattern with the price that the
$U(1)_\zeta$ charges of the Standard Model states 
do not satisfy the $E_6$ embedding.
It turns out that the $E_6$ embedding is necessary
for unified gauge couplings to agree with the low 
energy values of $\sin^2\theta_W(M_Z)$ and $\alpha_s(M_Z)$
\cite{viraf}. Construction of string models that admit the 
$E_6$ charges of the Standard Model states,
while maintaining $U(1)_\zeta$ as an anomaly free symmetry
was discussed in ref. \cite{afm}. The basic element 
of the proposed construction is to keep the 
massless chiral states in complete 27 representations
of $E_6$, while the $E_6$ symmetry is broken at the string
level and is not manifest in the string vacuum. 
In ref. \cite{fr} a PS heterotic--string derived model with 
anomaly free $U(1)_\zeta$ was obtained by using
the classification methodology developed in ref. 
\cite{classification}, and exploiting the spinor--vector
duality that was discovered in ref. \cite{svd}. 
The key ingredient is that the model of ref. \cite{fr}
is self--dual under the exchange of the total number 
of spinorial $16\oplus\overline{16}$ and vectorial 
$10$ representations of $SO(10)$. This is the same condition
that one has if the $SO(10)\times U(1)_\zeta$ symmetry
is enhanced to $E_6$. However, in the model of ref. 
\cite{fr} this is not the case, {\it i.e} the
$SO(10)$ symmetry is not enhanced to $E_6$. 
This is possible in the free fermionic model 
if the different 16 and 10+1 states,
that would make a complete 27 of $E_6$, are obtained 
from different fixed points of the underlying
$Z_2\times Z_2$ orbifold \cite{fr}. 

In the free fermionic SLM, PS and LRS models the weak hypercharge 
is given by\footnote{
$U(1)_C=3/2 U(1)_{B-L}$ and $U(1)_L= 2 U(1)_{T_{3_R}}$
are used in free fermionic models and will also be used below.}

\beq
U(1)_Y~=~ {1\over 2} U(1)_{B-L} + U(1)_{T_{3_R}}~,
\eeq
where $B-L$ is Baryon minus Lepton number and 
$T_{3_R}$ is the diagonal generator of $SU(2)_R$. 
The $SO(10)$ orthogonal combination is given by 
\beq
U(1)_{Z^\prime}~=~ {1\over 2} U(1)_{B-L} - {2\over 3} U(1)_{T_{3_R}}~\in~
SO(10).
\label{ubminusl}
\eeq
The VEV of the Higgs field with the quantum charges of the 
right--handed neutrino leaves unbroken the $U(1)_{{\cal Z}^\prime}$ combination,
\beq
U(1)_{{\cal Z}^\prime} ~=~
{3\over {10}} U(1)_{B-L} -{2\over5} U(1)_{T_{3_R}} - {1\over 5} U(1)_\zeta
~\notin~ SO(10),
\label{uzpwuzeta}
\eeq
that may remain unbroken down to low scales only if $U(1)_\zeta$ is 
anomaly free. 

\subsection{Observable non--universal $U(1)$s }

In addition to the family universal $U(1)$ symmetries in the observable 
$E_8$ gauge group, the string models contain two additional $U(1)$ symmetries
that are combinations of $U(1)_{1,2,3}$ and are orthogonal to $U(1)_\zeta$.
These are family non--universal and therefore must be heavier 
than roughly 30 TeV 
due to Flavour Changing Neutral Current (FCNC) constraints \cite{cfnv}. 
Additional observable $U(1)_{4,5,6}$ symmetries may arise from complexification 
of real fermions as discussed above. One combination of those may be 
family universal while the other two are not. In ref. \cite{pati}
it was proposed that the family universal anomaly free 
combination of $U(1)_{1,2,3,4,5,6}$ in the model of ref. \cite{slm}
plays a role in adequately suppressing proton decay mediating operators, 
as well as allowing for suppression of left--handed neutrino masses 
via the seesaw mechanism. However, it was shown in ref. \cite{plb499}
that the $U(1)$ discussed in ref. \cite{pati} must in fact be 
broken near the string scale. 
This is expected as this $U(1)$ symmetry is a combination 
of $U(1)_\zeta$, which is anomalous, with the family universal 
combination of $U(1)_{4,5,6}$.
In ref. \cite{thor} it was shown that two of the anomaly free non--universal 
combinations may similarly, adequately suppress proton decay and generate
small neutrinos via a seesaw mechanism. As discussed above 
they must be broken above the DecaTev scale. The additional combinations
of $U(1)_{1,2,3,4,5,6}$, 
aside from $U(1)_\zeta$, will not be considered further here. 

\subsection{Hidden sector $U(1)$s}

In addition to the $U(1)$ symmetries that arise in the observable sector, 
the string models may contain $U(1)_h$ symmetries that arise from 
the hidden $E_8$ gauge group. Such $U(1)_h$ symmetries may mix
with the weak hypercharge via kinetic mixing \cite{kineticmixing} 
provided that there exist light states in the spectrum that 
are charged under both $U(1)_Y$ and under the hidden
sector $U(1)_h$ factor. Depending on the details of the spectrum 
kinetic mixing may then arise from one--loop radiative corrections
\cite{kineticmixing} and is proportional to ${\rm Tr}Q_YQ_h$. 

The existence of hidden $U(1)_h$
symmetries in semi--realistic heterotic--string models is
highly model dependent, but there are some generic properties that may 
be highlighted. The PS class of models typically do not contain 
$U(1)$ factors in the hidden sector. The reason being that the PS 
models utilise only periodic/antiperiodic boundary conditions, and that
the set of basis vectors that generate a PS model typically contain a single 
$SO(10)$ breaking vector. 

The FSU5 models utilise rational boundary conditions, which break 
$SO(2n)$ symmetries
into $SU(n)\times U(1)$. Provided that the hidden sector 
gauge symmetry is not 
enhanced, the hidden sector may contain unbroken $U(1)$ 
factors. In the FSU5 model
of ref. \cite{fsu5} the hidden sector gauge group is enhanced 
and this model does
not have any hidden sector $U(1)$ factors. In the FSU5 models 
that were classified 
in ref. \cite{fsu5classi} all the hidden sector gauge group 
enhancements are projected 
out and therefore these FSU5 models do contain two hidden 
$U(1)$ symmetries.

The SLM \cite{fny, slm, fmt, cfmt} and LRS \cite{lrs}
models utilise two basis vectors that break the $SO(10)$ symmetry. 
These models generically contain several hidden sector $U(1)$ factors,
irrespective of whether 
the hidden sector symmetry is enhanced or not.

We now turn to discuss the matter states appearing in the models and 
the feasibility of kinetic mixing. Before getting into specific 
$SO(10)$ subgroups several broad observations can be made. 
All the models that we discuss have $N=1$ space--time supersymmetry, 
but the general properties that we extract are also applicable 
in tachyon free non--supersymmetric vacua \cite{aafs}.
The first division of the matter sectors is into those that preserve 
$N=4$, and those that preserve $N=2$, spacetime
supersymmetry. In the discussion of kinetic mixing
it is sufficient to focus on the $N=2$ sectors. These sectors are 
obtained from combinations of the basis vectors $b_{1,2,3}$ with 
the other basis vectors. The basis vectors $b_{1,2,3}$ in the NAHE--based 
models produce spinorial $SO(10)$ representations that are neutral under the 
hidden sector. The sectors $b_i+2\gamma$ produce states that transform 
as vector representations of the hidden sector gauge group, 
and are singlets of the $SO(10)$ subgroup. 
States that transform in the 10 vector representation of $SO(10)$ 
are neutral under the hidden sector gauge group. All the sectors
discussed thus far therefore cannot give rise to kinetic 
mixing with the weak hypercharge because they are not charged
with respect to both $U(1)_Y$ and $U(1)_h$. 

States that can induce kinetic mixing in free fermionic models 
can therefore only arise from sectors that break the $SO(10)$
symmetry. These sectors arise in combinations of the 
basis vectors $b_{1,2,3}$ with the $SO(10)$ basis vectors
$\alpha, \beta, \gamma$. Here we can further divide into 
sectors that break the $SO(10)$ symmetry to the PS or 
FSU5 subgroups. We will focus here on the examples of the
FSU5 and SLM models. In the case of the FSU5 models all 
$SO(10)$ breaking sectors contain states that carry 
fractional electric charge. The states may transform 
as singlets or fiveplets of $SU(5)$ and both type of states
will carry fractional electric charge. These states must therefore
be decoupled from the massless spectrum \cite{exoslm}, or confined
\cite{fsu5,fsu5classi}, at a high scale
and cannot generate sizable kinetic mixing. 

The SLM models contain a richer variety of $SO(10)$ breaking 
sectors, that can be divided according to the surviving $SO(10)$ 
subgroup, which can be $SU(5)\times U(1)$, $SO(6)\times SO(4)$ 
or $SU(3)\times SU(2)\times U(1)^2$ \cite{exoslm}. The first two cases
produce states with fractional electric charge, which
must be either decoupled or confined \cite{exoslm}. The last category of states
produces states that carry standard charges under the Standard Model gauge group
but carry non--standard $SO(10)$ charges under $U(1)_{Z^\prime}$.
One type of states in these sectors are neutral under the weak hypercharge
and therefore cannot generate kinetic mixing. The other type of states arising 
in these sectors are states that transform as 
$3$, ${\bar 3}$ and $2$, ${\bar 2}$
of the observable $SU(3)$ and $SU(2)$ groups, 
respectively, and carry the standard
Standard Model charge under $U(1)_Y$. These states interact
via the strong and 
electroweak interactions,
and therefore cannot remain light to the required scale to
produce sizable mixing \cite{kineticmixing}.
We conclude that kinetic mixing of a hidden sector $U(1)_h$ with $U(1)_Y$
is not viable in free fermionic models.  

For concreteness we can elaborate on this structure in some of the specific
heterotic--string standard--like models in the literature. For instance the 
model of ref. \cite{fmt}, which is given by the NAHE--set of basis vectors
plus the basis vectors $\{b_4, \beta, \gamma\}$ in eq. (3.2) of \cite{fmt}. 
In this model the observable and hidden sector
gauge symmetries are given by
\beqn
{\rm Observable}~& : &~~~~~~~~~SU(3)\times SU(2)\times U(1)_C\times U(1)_L 
                                               \times U(1)_{1,2,3,4,5,6}
\label{obsfmt}\\
{\rm Hidden}~& : &~~~~~~~~~SU(6)^2\times U(1)_7\times U(1)_8
\label{hiddenfmt}
\eeqn
The entire spectrum of the model is given in 
ref. \cite{fmt}. The sectors 
$b_i$, 
$b_i+2\gamma$,
$b_4+2\gamma$, 
${\bf 1}+b_1+b_2+b_3+b_4+2\gamma$
and the NS sector, where 
$i=1,2,3$,
produce states that are charged with respect to either 
$U(1)_Y$ or $U(1)_h$ but not with respect to both.
The states in the sectors 
${\bf 1}+b_j+b_k+2\gamma$, 
$j\ne k=1,2,3$,
are neutral with respect to both $U(1)_Y$ and $U(1)_h$. 
The sectors 
${\bf 1} + b_4 +\beta+2\gamma$,
${\bf 1} + b_4 + \beta$,
${\bf 1} + b_1 +b_2 + b_4 \pm\gamma$,
${\bf 1} + b_1 +b_2 + b_3 +\beta+2\gamma$,
$\pm\gamma$,
$b_1+b_3\pm\gamma$,
${\bf 1} +b_4 + \pm\gamma$,
$b_3 +b_4 \pm\gamma$ 
and 
$b_1 +b_2 +b_3 +b_4 \pm\gamma$
produce vector--like states that carry fractional $\pm1/2$ charge
and must be decoupled or confined at a high scale \cite{exoslm}. 
The sectors
${\bf 1} + b_3 + b_4 + \beta + \pm\gamma$
and 
${\bf 1} + b_2 + b_4 + \beta + 2\gamma$
produce exotic states that are neutral under $U(1)_Y$ and 
charged under $U(1)_h$. 
Similar structure of the spectrum with respect to states
that can potentially mix between $U(1)_Y$ and $U(1)_h$
arises in the models of refs. \cite{fny, slm, cfmt}. 
We conclude that kinetic mixing of $U(1)_Y$ and $U(1)_h$ 
in these free fermionic models is not viable.

\section{Light $U(1)$s}\label{lightu1s}

In this section we consider the possibility that an extra $U(1)$ symmetry is
left--unbroken in the heterotic--string vacuum; the
phenomenological constraints; and the distinctions
between the different models. The four cases that we discuss are: 
(i) the $U(1)_{Z^\prime}$ in eq. (\ref{ubminusl}); 
(ii) the $U(1)_{{\cal Z}^\prime}$ in eq. (\ref{ubminusl}); 
(iii) the non--Abelian left--right symmetric extension $SU(2)_R\times U(1)_C$;
(iv) the PS models. 
For completeness we also mention two additional cases: 
(v) the $SU(4)\times SU(2)\times U(1)_C$ models;
(vi) the leptophobic $Z^\prime$ and custodial $SU(2)$ models.

The main phenomenological constraints are 
with respect to proton stability and the suppression of 
left--handed neutrino masses. Specifically, the simultaneous 
accommodation of both constraints is problematic. The reason being that 
while proton stability favours baryon number conservation, 
suppression of neutrino masses demands that lepton number 
is  violated. In the free fermionic models baryon minus 
lepton number is gauged and therefore breaking lepton
number implies that baryon number is broken as well, 
giving rise to dimension four proton decay mediating 
operators from non renormalisable operators \cite{ps}, 
\beq
QLD{\cal N}{\phi^n} ~~~~~~~~~~~~~udd{\cal N}{\phi^n}
\label{deqfourop}
\eeq
where $\phi^n$ is a string of states that get a vacuum expectation 
value of the order of the string scale, whereas ${\cal N}$ and ${\bar{\cal N}}$
are the components of the heavy Higgs fields that break $U(1)_{Z^\prime}$. 
The operators in eq. (\ref{deqfourop})
arise from the $16^4$ operator of $SO(10)$ and therefore
arise in any of the string models discussed above. It is noted from 
(\ref{deqfourop}) that the magnitude of the proton decay mediating operators 
is proportional to the scale of $U(1)_{Z^\prime}$ breaking. This is a general 
feature of the $SO(10)$ based free fermionic models. 

On the other hand the structure of the neutrino mass matrix is also 
quite generic in these models. In term of component fields, the terms in the 
superpotential that generate the neutrino mass matrix are (see {\it e.g.}
\cite{tauneutrinomass}),
\beq
L_iN_j{\bar h}~~~,~~~N_i{\bar {\cal N}}\phi_j~~~,~~~\phi_i\phi_j\phi_k~,
\label{supterms}
\eeq
where $L_i$, $N_i$ and $\phi_i$, with $i,j,k=1,2,3$ 
are the lepton doublets; the right--handed neutrinos; and three 
$SO(10)$ singlet fields, respectively;
${\bar h}$ is the electroweak Higgs doublet and ${\bar {\cal N}}$ is the
component of the heavy Higgs field that breaks $U(1)_{Z^\prime}$. 
All these states exist in the spectra of the string models, possibly
as components of larger representation in, {\it e.g.}, the FSU5 models. 
The neutrino seesaw mass matrix takes the generic form 
\begin{equation}
{\left(
\begin{matrix}
                 {\nu_i}, &{N_i}, &{\phi_i}
\end{matrix}
   \right)}
  {\left(
\begin{matrix}
             0   &       (M_{_D})_{ij}           &             0                    \\
     (M_{_D})_{ij}&          0                   &     \langle{\bar {\cal N}}\rangle_{ij} \\
             0   &\langle{\bar {\cal N}}\rangle_{ij} &    \langle\phi\rangle_{ij}              \\
\end{matrix}
   \right)}
  {\left(
\begin{matrix}
                 {\nu_j}  \cr
                 {N_j}\cr
                 {\phi_j} \cr
\end{matrix}
   \right)},
\label{nmm}
\end{equation}
where $M_{_D}$ is the Dirac mass matrix arising from the first term in eq. 
(\ref{supterms}). Due to the underlying $SO(10)$ symmetry the Dirac mass 
matrix is proportional to the up--quark matrix \cite{tauneutrinomass}. 
At the cubic level of the superpotential the symmetry dictates the 
equality of the top quark and tau neutrino Yukawa couplings. 
Hence, for the tau neutrino we have that $M_{_D}=k M_{\rm top}$, where $k$
is a renormalisation factor due to RGE evolution. Taking the mass matrices 
to be diagonal the mass eigenstates are primarily $\nu_i$, $N_i$ and $\phi_i$ 
with negligible mixing and with the eigenvalues 
\beq
m_{\nu_j} \sim  
\left(
{{k M^j_u} \over {\langle{\bar {\cal N}}\rangle  }    }
\right)^2 \langle{\phi}\rangle~~~,~~~
\qquad m_{N_j},m_{\phi} \sim \langle{\bar {\cal N}}\rangle~.
\label{neutrinomasseigen}
\eeq
Therefore, the left--handed neutrino masses are inversely 
proportional to the 
square of the $U(1)_{Z^\prime}$ breaking scale and to the VEV of the 
$SO(10)$ singlet 
field $\phi$. This structure is generic in this class of models and 
the question 
is what is required in order to accommodate the left--handed neutrino 
masses in the 
different scenarios. Detailed studies of the neutrino masses in 
free fermionic models 
were performed in \cite{sseasaw}. 
Here we are only interested in the qualitative features. 
We can then consider several cases. 

\subsection{Case i: low $B-L$ breaking scale}\label{case1}

In this case the spectrum contains the MSSM states plus the 
right--handed neutrinos;
a pair of Higgs doublets that break the electroweak symmetry and a pair of
Higgs singlets that break the $U(1)_{Z^\prime}$ symmetry \cite{zpbminusl}. 
The full spectrum is
displayed in table \ref{bminuslmodel}.  

\begin{table}[!h]
\noindent 
{\small
\begin{center}
{\tabulinesep=1.2mm
\begin{tabu}{|l|cc|c|c|c|}
\hline
Field &$\hphantom{\times}SU(3)_C$&$\times SU(2)_L $
&${U(1)}_{Y}$&${U(1)}_{Z^\prime}$  \\
%\hline
\hline
$Q_L^i$&    $3$       &  $2$ &  $+\frac{1}{6}$   & $+\frac{1}{2}$   \\
$u_L^i$&    ${\bar3}$ &  $1$ &  $-\frac{2}{3}$   & $+\frac{1}{2}$  \\
$d_L^i$&    ${\bar3}$ &  $1$ &  $+\frac{1}{3}$   & $-\frac{3}{2}$  \\
$e_L^i$&    $1$       &  $1$ &  $+1          $   & $+\frac{1}{2}$  \\
$L_L^i$&    $1$       &  $2$ &  $-\frac{1}{2}$   & $-\frac{3}{2}$  \\
$N_L^i$&    $1$       &  $1$ &  ~~$0$            & $+\frac{5}{2}$ \\
\hline
$h$         & $1$       & $2$ &  $-\frac{1}{2}$  &   $+1$   \\
${\bar h}$  & $1$       & $2$ &  $+\frac{1}{2}$  &   $-1$  \\
\hline
$\phi^i$       & $1$       & $1$ &  ~~$0$  &  ~~$0$   \\
\hline
\hline
${\cal N}$         & $1$       & $1$ &   ~~$0$   &  $+\frac{5}{2}$     \\
${\bar {\cal N}}$  & $1$       & $1$ &   ~~$0$   &  $-\frac{5}{2}$    \\
\hline
\end{tabu}}
\end{center}
}
\caption{\label{bminuslmodel}
\it
Spectrum and
$SU(3)_C\times SU(2)_L\times U(1)_{Y}\times U(1)_{Z^\prime}$ 
quantum numbers, with $i=1,2,3$ for the three light 
generations. The charges are displayed in the 
normalisation used in free fermionic 
heterotic--string models. }
\end{table}

Taking $m_t\sim 173$GeV; $k\sim1/3$; 
$\langle {\bar {\cal N}}\rangle \sim 3$ TeV
we note that to accommodate a tau neutrino mass below 1eV we need 
$\langle \phi \rangle \sim 1$keV. While not impossible, it requires 
the introduction of a new scale, which may %not 
be ad hoc from the string
model building perspective \cite{sseasaw}.

\subsection{Case ii: high $B-L$ breaking scale}\label{case2}

In this case we assume that the VEV of ${\cal N}$ is high, or intermediate. 
Furthermore, we may assume that 
$\langle \phi \rangle \sim 100$GeV, {\it i.e.} that this VEV is associated 
with electroweak symmetry breaking. Then 
taking $\langle {\bar {\cal N}}\rangle \sim 10^{17}$GeV
gives $m_{\nu_\tau}\sim 10^{-20}$GeV. Breaking $U(1)_{Z^\prime}$
at the high scale therefore
naturally produces light neutrino masses,
with the scale of $\langle\phi\rangle$ 
being associated with the electroweak breaking scale.
In this case the combination 
$U(1)_{{\cal Z}^\prime}$ 
in eq. (\ref{uzpwuzeta}) remains unbroken. This is possible if and only if 
$U(1)_\zeta$ is anomaly free. As discussed above this necessitates that the 
chiral states form complete 27 representations of $E_6$.
However, the normalisation 
of $U(1)_{{\cal Z}^\prime}$ may differ from the standard $E_6$
normalisation, similar 
to the discussion in relation to the normalisation of the
weak hypercharge \cite{u1normalisation}.
The spectrum of the string inspired model that may keep
$U(1)_{{\cal Z}^\prime}$ unbroken down to the TeV scale is shown in table
\ref{table27rot}. The effective dimension four operators induced
from eq. (\ref{deqfourop}) are not invariant under $U(1)_{{\cal Z}^\prime}$. 
Hence, the dimension four proton decay operators are 
suppressed as in the case with a low $U(1)_{Z^\prime}$ of section
\ref{case1}. The caveat is that the spectrum contains
leptoquark representations
that arise from the $SO(10)$ vectorial 10 representation, 
and may mediate rapid proton decay
\cite{ps}. Additional discrete symmetries are required to
guarantee adequate suppression
of the dangerous operators. This issue arises generically in string inspired 
${\cal Z}^\prime$ models with an underlying $E_6$ symmetry \cite{e6zprime},
{\it i.e.} in all 
models in which $U(1)_\zeta$ forms part of the low scale $Z^\prime$. We note that 
this is not a problem in the model of section \ref{case1} because there 
$U(1)_\zeta$ does not enter into the combination of the low scale $Z^\prime$. 
We note again that the root of the problem is the conflict between adequately
suppressing proton decay mediating operators,
which favours a low scale $U(1)_{Z^\prime}$
and the constraint of left--handed neutrino masses, which works more naturally 
with $U(1)_{Z^\prime}$ being broken at a high scale.

\begin{table}[!h]
\noindent 
{\small
\begin{center}
{\tabulinesep=1.2mm
\begin{tabu}{|l|cc|c|c|c|}
\hline
Field &$\hphantom{\times}SU(3)_C$&$\times SU(2)_L $
&${U(1)}_{Y}$&${U(1)}_{Z^\prime}$&  \\
%\hline
\hline
$Q_L^i$&    $3$       &  $2$ &  $+\frac{1}{6}$   & $-\frac{2}{3}$ &  ~~  \\
$u_L^i$&    ${\bar3}$ &  $1$ &  $-\frac{2}{3}$   & $-\frac{2}{3}$ &  ~~  \\
$d_L^i$&    ${\bar3}$ &  $1$ &  $+\frac{1}{3}$   & $-\frac{4}{3}$ &  ~~  \\
$e_L^i$&    $1$       &  $1$ &  $+1          $   & $-\frac{2}{3}$ &  ~~  \\
$L_L^i$&    $1$       &  $2$ &  $-\frac{1}{2}$   & $-\frac{4}{3}$ &  ~~  \\
$N_L^i$&    $1$       &  $1$ &  ~~$0$            & ~~$0$ &  ~~  \\
\hline
$D^i$       & $3$     & $1$ & $-\frac{1}{3}$     & $+\frac{4}{3}$ & ~~    \\
${\bar D}^i$& ${\bar3}$ & $1$ &  $+\frac{1}{3}$  &   ~~$ 2$       &   ~~    \\
$H^i$       & $1$       & $2$ &  $-\frac{1}{2}$   &   ~~$2$       & ~~    \\
${\bar H}^i$& $1$       & $2$ &  $+\frac{1}{2}$   &   $+\frac{4}{3}$ &  ~~  \\
\hline
$S^i$       & $1$       & $1$ &  ~~$0$  &  $-\frac{10}{3}$       &   ~~   \\
\hline
$h$         & $1$       & $2$ &  $-\frac{1}{2}$  &  $-\frac{4}{3}$  & ~~    \\
${\bar h}$  & $1$       & $2$ &  $+\frac{1}{2}$  &  $+\frac{4}{3}$  & ~~    \\
\hline
$\phi^i$       & $1$       & $1$ &  ~~$0$  &  ~~$0$       &   ~~   \\
\hline
\end{tabu}}
\end{center}
}
\caption{\label{table27rot}
\it
Spectrum and
$SU(3)_C\times SU(2)_L\times U(1)_{Y}\times U(1)_{{\cal Z}^\prime}$ 
quantum numbers, with $i=1,2,3$ for the three light 
generations. The charges are displayed in the 
normalisation used in free fermionic 
heterotic--string models. }
\end{table}

\subsection{Case iii: Low scale left--right symmetric models}\label{case3}

In the LRS models with a low $U(1)_{Z^\prime}$ breaking the Standard model states 
are organised in representations of the low scale gauge symmetry, 
\beq
SU(3)_C\times U(1)_C\times SU(2)_L \times SU(2)_R
\label{lrsgs}
\eeq
In this case the $U(1)_{Z^\prime}$ combination is identical to the combination
given in eq. (\ref{ubminusl}). However, in this case additional 
$W^\prime$ vector bosons
arise. The spectrum of the model is shown in table 
\ref{leftrightsymmetricmodel}.
The dimension four
proton decay mediating operators arise from the terms 
\beq
Q_LL_LQ_R{\cal L}_R\phi^n ~~~~{\rm and} ~~~~Q_RQ_RQ_R{\cal L}_R\phi^n
\label{lrsdeq4}
\eeq

\begin{table}[!h]
\noindent 
{\small
\begin{center}
{\tabulinesep=1.2mm
\begin{tabu}{|l|ccc|c|c|}
\hline
Field &$\hphantom{\times}SU(3)_C$&$\times SU(2)_L $ & $SU(2)_R $
&${U(1)}_{C}$&${U(1)}_{\zeta}$ \\
%\hline
\hline
$Q_L^i$  & $3$       & $2$ & $1$ & $+\frac{1}{2}$   & $-\frac{1}{2}$   \\
$Q_R^i$  & ${\bar3}$ & $1$ & $2$ & $-\frac{1}{2}$   & $+\frac{1}{2}$  \\
$L_L^i$  & $1$       & $2$ &  $1$ & $-\frac{3}{2}$   & $-\frac{1}{2}$  \\
$L_R^i$  & $1$       & $1$ &  $2$ & $+\frac{3}{2}$   & $+\frac{1}{2}$ \\
\hline
${\cal L}_R$  & $1$       & $1$ &  $2$ & $+\frac{3}{2}$   & $+\frac{1}{2}$ \\
${\bar {\cal L}}_R$ & $1$ & $1$ &  $2$ & $-\frac{3}{2}$   & $-\frac{1}{2}$ \\
\hline
$h$      & $1$      & $2$ &  $2$ & ~~$0$            &  $0$      \\
\hline
$\phi^i$ & $1$     & $1$ & ~~$0$ &  ~~$0$           &  $0$           \\
\hline
\end{tabu}}
\end{center}
}
\caption{\label{leftrightsymmetricmodel}
\it
Spectrum and
$SU(3)_C\times U(1)_C\times SU(2)_L\times SU(2)_{R}\times U(1)_{\zeta}$ 
quantum numbers, with $i=1,2,3$ for the three light 
generations. The charges are displayed in the 
normalisation used in free fermionic 
heterotic--string models. }
\end{table}

We note that as $U(1)_{{\cal Z}^\prime}$ is broken at a high scale in this
scenario both terms can be generated without the adequate suppression discussed 
in refs. \cite{plb499, cfgejp}. However, as in section \ref{case1}
they are adequately suppressed due to the fact that 
$U(1)_{Z^\prime}$ is broken at a low scale, {\it i.e.} $U(1)_{B-L}$ 
is gauged down to low scales. The left--right symmetric models only
require the existence of the right--handed neutrinos in the spectrum, 
but not the states from the vectorial 10 representation of $SO(10)$. 
However, similar to the case in section \ref{case1}, a Yukawa coupling
of the Dirac mass
term for the tau neutrino is of the order of the top quark mass
and we have to assume the existence of a scale of the order of 
1KeV as in section \ref{case1}. This is the case in the LRS string
derived model of ref. \cite{lrs}.
An alternative possibility that may be contemplated
is that only the mass term of the 
top quark is generated at cubic order of the superpotential,
whereas the coupling of the tau neutrino to the same 
Higgs bi--doublet is obtained from higher order nonrenormalisable 
terms. In this case the relation between the top quark and tau 
neutrino Dirac mass term can be avoided. The tau neutrino 
Yukawa coupling is equal to that of the tau lepton, where 
the two 
relevant mass terms are $\lambda_t Q_L^tQ_R^th$ 
and $\lambda_\tau L_L^tL_R^th$. Therefore
up to running effects the tau neutrino Dirac mass term will
be of the order of the tau lepton mass. 
Taking $m_\tau\sim 1.776$GeV
and assuming a seesaw scale of the order of 10 TeV requires
$\langle\phi\rangle\sim 10$MeV. 
%
%
%field theory constructions is that the 
%Higgs bi--doublet that couples to the quarks is distinct from 
%the Higgs bi-doublet that couples to the leptons, This is not the case in the 
%string derived model of ref. \cite{lrs}, and would require 
%two bi--doublets at low scales, which may be in conflict with 
%gauge coupling unification. 
An interesting observation is that the string derived
left--right symmetric heterotic--string models
allow for the nonrenormalisable terms
\beq
L_LL_L{L}_R{L}_R, 
\eeq
due to 
the $U(1)_\zeta$ charges 
in these models, as displayed in table \ref{leftrightsymmetricmodel}.
Assuming that the electrically neutral scalar component of $L_R$ gets 
a VEV of the
order of $3$ TeV, we get a Majorana mass term for the left--handed
neutrino of order $\langle{\tilde N}\rangle^2/M_S$, where $M_S$ is a scale 
of the order of the string scale, $M_S\sim 5\times 10^{17}{\rm GeV}$. 
The effective Majorana mass for the left--handed neutrinos is then
of order $10^{-1}$eV. This possibility enables the breaking of $SU(2)_R$ 
without the additional Higgs fields ${\cal L}_R$ and ${\bar{\cal L}}_R$, 
which is advantageous for gauge coupling unification \cite{gcu}. 

\subsection{Case iv: Low scale Pati--Salam models}\label{case4}

In the PS models the low energy effective gauge symmetry below the
string scale is the $SO(10)$ subgroup $SO(6)\times SO(4)$. 
The possibility of the Pati--Salam symmetry \cite{patisalam}
at the 
TeV scale was discussed in ref. \cite{patisalam,volkas}. 
Similarly to the case of $U(1)_{Z^\prime}$ and the left--right 
symmetry models anomaly cancellation only requires the addition of
three right--handed neutrinos to the Standard Model states. 
The vector bosons in this model do not generate Proton
decay via dimension six operators. A low scale breaking of the 
PS symmetry can therefore be considered. The spectrum 
of the model is shown in table \ref{PSmodel}. 
\begin{table}[!h]
\noindent 
{\small
\begin{center}
{\tabulinesep=1.2mm
\begin{tabu}{|l|ccc|}
\hline
Field &$\hphantom{\times}SU(4)_C$&$\times SU(2)_L $ & $SU(2)_R $ \\
%\hline
\hline
${\cal Q}_L^i$  & $4$       & $2$ & $1$       \\
${\cal Q}_R^i$  & ${\bar4}$ & $1$ & $2$   \\
\hline
${\cal H}$  & ${\bar4}$ & $1$ & $2$   \\
${\bar{\cal H}}$  & ${4}$ & $1$ & $2$   \\
$D$      & $6$      & $1$ & $1$      \\
$h$      & $1$      & $2$ &  $2$             \\
\hline
$\phi^i$ & $1$     & $1$ & ~~$0$              \\
\hline
\end{tabu}}
\end{center}
}
\caption{\label{PSmodel}
\it
Spectrum and
$SU(4)_C\times SU(2)_L\times SU(2)_{R}$ 
quantum numbers, with $i=1,2,3$ for the three light 
generations. The charges are displayed in the 
normalisation used in free fermionic 
heterotic--string models. }
\end{table}

A low scale breaking of the PS symmetry may be obtained via the VEV
of the neutral scalar component in a $({\bar 4},1,2)$ representation, 
whereas a high scale breaking requires an additional pair
of heavy Higgs fields, ${\bar {\cal H}}\oplus{\cal H}=
({\bar 4}, 1,2)_{\cal H}~\oplus~({ 4}, 1,2)_{\cal H}$, 
%in addition to the states in table \ref{PSmodel}, 
to break the symmetry along supersymmetric
flat directions. The dimension four operators are induced from the 
quartic order terms ${\cal Q}_L{\cal Q}_L{\cal Q}_R{\cal Q}_R$
and ${\cal Q}_R{\cal Q}_R{\cal Q}_R{\cal Q}_R$. With a low 
breaking of $SU(2)_R$ these operators are sufficiently suppressed. 
The PS model with a high scale breaking include the $(6,1,1)$ 
representation to generate mass to the coloured states 
of the heavy Higgs states, via the couplings
${\bar {\cal H}}{\bar {\cal H}}D+{\cal H}{\cal H}D$. 
With a low scale breaking these states are not required 
because an additional pair of heavy Higgs states is not 
required as the breaking can be implemented along a non 
flat direction. In this model suppression of left--handed 
neutrino masses may be obtained by the generations of VEVs of the
order of 1keV, similar to the discussion in section 
\ref{case1}, or may be generated from the quartic order coupling 
${\cal Q}_R {\cal Q}_R {\cal Q}_R {\cal Q}_R$ as in section
\ref{case3}. We note that the mass structure of the extra vector states
in this PS scenario requires elaborate analysis, with the possibility
that the charged $W^\prime$s are relatively light, whereas the 
neutral $U(1)_{Z^\prime}$ is comparatively heavy, as is the
case in the Standard Model. These considerations raise the 
prospect that there will be a need to probe the DecaTev scale and above. 

\subsection{Case v: Low scale $SU(4)\times SU(2)\times U(1)_L$ 
models}\label{case5}

For completeness we comment on the case with $SO(10)$ broken to the 
$SU(4)\times SU(2)\times U(1)_L$ model\footnote{we note that 
$U(1)_L= 2 U(1)_{T_{3_R}}$, where $T_{3_R}$ is the diagonal generator
of $SU(2)_R$.}. This model was considered in ref. \cite{wise}
as a field theory extension of the Standard Model. The field theory
model considered in ref. \cite{wise} utilise Higgs field in the 
$(15,2,1)$ representation of $SU(4)\times SU(2)\times U(1)_L$, 
to avoid the relation between the Dirac mass terms of the top quark
and the tau neutrino. The string models do not contain such representations
and therefore the only available route to satisfy the neutrino
mass constraints is to assume $\langle\phi\rangle\sim1$keV. 
The $SU(4)\times SU(2)\times U(1)$ choice for the $SO(10)$ subgroup 
of the string model is attractive because it admits both the 
doublet--triplet splitting mechanism \cite{dts} as well as the 
doublet--doublet splitting mechanism \cite{su421}.
However, as discussed above, while a field theory model consistent 
with the phenomenological constraints can be constructed \cite{su421}, 
it was shown in ref. \cite{su421} that such string models
are not viable because it is not possible to form complete families.
This demonstrates that the string constructions are more restrictive 
than the field theory constructions. This is anticipated as 
the string framework consistently incorporates gravity into the construction.
An alternative method to produce $SU(4)\times SU(2)\times U(1)$ 
three generation
vacua is by enhancement of the NS--gauge group from additional 
sectors \cite{custsu2, leptozprime}. 
%However, in such models
%the weak--hypercharge may or may not have an $SO(10)$ embedding.
 
\subsection{Case vi: Leptophobic $Z^\prime$ and custodial $SU(2)$s}
\label{case6}

Finally, we comment briefly on the possibility of generating leptophobic 
$Z^\prime$ 
\cite{leptozprime} and custodial $SU(2)$ symmetries
\cite{custsu2} in the free fermionic
heterotic--string models. As mentioned in section \ref{case5}
the gauge group arising from the NS--sector may be enhanced
by space--time vector bosons that are obtained from additional 
sectors in the additive group. Examples of such three generation string
models were presented in refs. \cite{custsu2, leptozprime, lrs}.
In these models the three generations still arise from the sectors 
$b_{1,2,3}$ and hence descend from the spinorial $16$ representations of
$SO(10)$, but they transform in representations of the enhanced 
gauge symmetry. 
Leptophobic $U(1)$s are obtained when $U(1)_{B-L}$ combines with 
the a universal combination of the horizontal flavour symmetries
to cancel out the lepton number and produce a gauged $U(1)_B$
\cite{leptozprime}. 
We note that in the custodial $SU(2)$ model only the lepton transforms
as doublets of $SU(2)_C$ \cite{custsu2}. Hence, the model will
have distinct signature compared to the LRS models of section
\ref{case3}. 
Namely, the additional $W^\prime$ vector bosons couple to leptons
but not to the hadrons, whereas a leptophobic $Z^\prime$ \cite{leptozprime}
couples to hadrons but not to leptons. 

%%%%%%%%
%%%%%%%%
\section{Prospects at the LHC}\label{lhcprop} 
 
In this section we illustrate LHC prospects for a hypothetical phenomenological scenario 
of a low scale heterotic-string derived $Z^\prime$, based on the 
high B-L breaking scale model of section~\ref{case2}. 
In particular, we show the LHC 8 TeV Drell-Yan (DY) invariant mass distribution at the 
{\it next-to-next-to} leading order (NNLO) in the QCD strong coupling constant 
(${\cal O}(\alpha_s^2)$)~\cite{Hamberg:1990np}, for the production
of a $Z'$ with mass $M_{Z'} = 3$ TeV. 

Recently, the ATLAS~\cite{atlas1} and CMS~\cite{cms} collaborations have published 
measurements of the DY differential cross section $\textrm{d}\sigma/\textrm{d}M$ 
in bins of dilepton invariant mass $M$ at center-of-mass energies $\sqrt{S}$ of 7 and 8 TeV.
In particular, $\textrm{d}\sigma/\textrm{d}M$ has been measured as a function of the 
invariant mass of dielectron and dimuon pairs, up to 2 TeV. 
These measurements are very precise in the mass region around the $Z_0$ peak, and 
no significant deviations from the SM prediction have been observed in the mass range explored. 
However, data at large invariant mass are still affected by large uncertainties 
due to systematical, statistical and luminosity errors.    

The uncertainty associated to scale variation of the NNLO QCD theory prediction 
amounts to a few percent, while 
that associated to the parton distribution functions (PDFs) luminosity is 
larger than 15\%, especially in the large invariant mass region. 
In this kinematic region PDFs are probed at large $x$, where they are in general not well constrained. 
This has a significant impact on the parton luminosity uncertainties 
as they are the major source of uncertainty and represent a limiting factor 
to obtain precise predictions for the production of high-mass dilepton resonances. 

LHC run-II will allow us to measure this and other differential observables
with higher precision in the high-mass region, and thus will 
confirm or rule out the existence of extra $Z'$s in the mass range of a few TeV's.

\subsection{Details of the calculation}

In this section we briefly describe the details of the 
calculation and the choice of the parameter space.
Electroweak corrections~\cite{Balossini:2006zz,Balossini:2008cs,Baur:2001ze,Zykunov:2005tc,CarloniCalame:2007cd,Arbuzov:2007db} 
are not included here, a more thorough analysis exploiting other differential observables~\cite{FEWZ1,FEWZ2} is left for future studies.

The theory is calculated by using an amended version of \textsc{CandiaDY}~\cite{cfgprd,Cafarella:2007tj},
a program that calculates the DY invariant mass distribution up to NNLO in QCD 
for a large variety of $Z'$ string derived models. 
The full spin correlations as well as the $\gamma^*/Z/Z'$ 
interference effects are included in this calculation.
The charge assignment is that of the high $B-L$ breaking scale model described in Sec.~\ref{case2} 
and is given in table \ref{table27rot}. Furthermore, we have chosen $\tan\beta=10$, the $Z'$ 
coupling constant $g_z$ equal to the hypercharge $g_Y$, and $M_{Z'}=$ 3 TeV.

The colour-averaged inclusive differential cross section is given by 
\ba
\frac{d\sigma}{dM^2}=\tau \sigma_{V}(M^2,M_V^2)
W_{V}(\tau,M^2)\hspace{1cm} \tau=\frac{M^2}{S},
\ea
where ($V=Z, Z^\prime$), and all the hadronic initial state information is contained in the
hadronic structure function which is defined as
\ba
W_{V}(\tau,M^2)=\sum_{i,j} \int_{0}^{1}dx_1 \int_0^1 dx_2 \int_{0}^{1}dx
\delta(\tau-x x_1 x_2)
{\cal L}_{i,j}^{V}(x_1,x_2,\mu_F^2)\Delta_{i,j}(x,M^2,\mu_F^2).
\nonumber\\
\ea
The contribution $W_{V}$ takes into account all the initial state emissions of real gluons and
all the virtual corrections, while $\sigma_{V}$ is the point-like cross section.
The parton luminosity ${\cal L}_{i,j}$ 
includes combinations of PDFs relative to the 
partonic structure of the initial state, while the hard scattering contributions, denoted by 
$\Delta_{i,j}(x,M^2,\mu_F^2)$, can be perturbatively expanded 
in terms of the strong coupling constant $\alpha_s(\mu_R^2)$
\ba
\Delta_{i,j}(x,M^2,\mu_F^2)=\sum_{n=0}^{\infty}
\alpha_s^n(\mu_R^2)\Delta^{(n)}_{i,j}(x,M^2,\mu_F,\mu_R^2)\,.
\ea
$\mu_F$ and $\mu_R$ are the factorization and renormalization scales respectively, 
while the invariant mass of the dilepton pair is denoted by $M$.

We strictly follow the notation introduced in Ref.~\cite{cfgprd} 
and here we briefly recall the main definitions.
The fermion-fermion-$Z^{\prime}$ interaction is given by
\ba
\sum_f z_f g_z \bar{f} \gamma^{\mu} f Z_{\mu}^{\prime},
\ea
where $f=e_{R}^{j},l_{L}^{j},u_{R}^{j},d_{R}^{j},q_{L}^{j}$ and
$q_L^{j}=(u_{L}^{j},d_{L}^{j})\,\,,l_L^{j}=(\nu_{L}^{j},e_{L}^{j})$.
The coefficients $z_{u},z_{d}$ are the charges of the right-handed up and
down quarks, respectively, while the $z_q$ coefficients are the charges
of the left-handed quarks. 

The masses of the neutral gauge bosons are parametrized in terms of the charges and 
vev's of the higgs sector as 
\ba
&&\varepsilon=\frac{\delta M^2_{Z 
Z^{\prime}}}{M^2_{Z^{\prime}}-M^2_{Z}}\nonumber\\
&&M_Z^2=\frac{g^2}{4 
\cos^2\theta_W}(v_{H_1}^2+v_{H_2}^2)\left[1+O(\varepsilon^2)\right]
\nonumber\\
&&M_{Z^{\prime}}^2=\frac{g_z^2}{4}(z_{H_1}^2 
v_{H_1}^2+z_{H_2}^2v_{H_2}^2+z_{\phi}^2 
v_{\phi}^2)\left[1+O(\varepsilon^2)\right]
\nonumber\\
&&\delta M^2_{Z Z^{\prime}}=-\frac{g g_z}{4\cos\theta_W}(z_{H_1}^2 
v_{H_1}^2+z_{H_2}^2v_{H_2}^2),
\ea
where $\varepsilon$ is defined as a perturbative parameter, and where $g=e/\sin\theta_W$ $g_Y= e/\cos\theta_W$.
The interaction Lagrangian for the quarks and the leptons is written as
\ba
&&{\mathcal{L}}_{int}=
\bar{Q}_{L}^{j}N^{Z}_{L}\gamma^{\mu}Q^{j}_{L} Z_{\mu}
+\bar{Q}_{L}^{j}N^{Z^{\prime}}_{L}\gamma^{\mu}Q^{j}_{L} Z^{\prime}_{\mu}
+\bar{u}_{R}^{j}N^{Z}_{u,R}\gamma^{\mu}u^{j}_{R} Z_{\mu}
\nonumber\\
&&\hspace{1cm}
+\bar{d}_{R}^{j}N^{Z}_{d,R}\gamma^{\mu}d^{j}_{R} Z_{\mu}
+\bar{u}_{R}^{j}N^{Z^{\prime}}_{u,R}\gamma^{\mu}u^{j}_{R} Z^{\prime}_{\mu}
+\bar{d}_{R}^{j}N^{Z^{\prime}}_{d,R}\gamma^{\mu}d^{j}_{R} Z^{\prime}_{\mu}
\nonumber\\
&&\hspace{1cm}
+\bar{Q}_{L}^{j}N^{\gamma}_{L}\gamma^{\mu}Q^{j}_{L} A_{\mu}
+\bar{u}_{R}^{j}N^{\gamma}_{u,R}\gamma^{\mu}u^{j}_{R} A_{\mu}
+\bar{d}_{R}^{j}N^{\gamma}_{d,R}\gamma^{\mu}d^{j}_{R} A_{\mu}
\nonumber\\
&&\hspace{1cm}
+\bar{l}_{L}^{j}N^{\gamma}_{L}\gamma^{\mu}l^{j}_{L} A_{\mu}
+\bar{e}_{R}^{j}N^{\gamma}_{e,R}\gamma^{\mu}e^{j}_{R} A_{\mu}
\nonumber\\
&&\hspace{1cm}
+\bar{l}_{L}^{j}N^{Z}_{L,lep}\gamma^{\mu}l^{j}_{L} Z_{\mu}
+\bar{l}_{L}^{j}N^{Z^{\prime}}_{L,lep}\gamma^{\mu}l^{j}_{L} Z^{\prime}_{\mu}
\nonumber\\
&&\hspace{1cm}
+\bar{e}_{R}^{j}N^{Z}_{e,R}\gamma^{\mu}e^{j}_{R} Z_{\mu}
+\bar{e}_{R}^{j}N^{Z^{\prime}}_{e,R}\gamma^{\mu}e^{j}_{R} Z^{\prime}_{\mu}\,,
\ea
where left-handed (L) and right-handed (R) couplings for the quarks are
\ba
&&N^{Z,j}_{L}=-i\left( g\cos\theta_W T^{L}_{3} -g_Y \sin\theta_W 
\frac{\hat{Y}^{L}}{2} + g_z \varepsilon\frac{\hat{z}^{L}}{2}\right)
\nonumber\\
&&N^{Z^{\prime},j}_{L}=-i\left(-g\cos\theta_W T^{L}_{3}\varepsilon
+g_Y\sin\theta_W  
\frac{\hat{Y}^{L}}{2}\varepsilon+g_z\frac{\hat{z}^{L}}{2}\right)
\nonumber\\
&&N^{Z}_{u,R}=-i\left(
-g_Y \sin\theta_W \frac{\hat{Y}^{u,R}}{2}+g_z 
\varepsilon\frac{\hat{z}^{u,R}}{2}\right)
\nonumber\\
&&N^{Z}_{d,R}=-i\left(
-g_Y \sin\theta_W \frac{\hat{Y}^{d,R}}{2}+g_z
\varepsilon\frac{\hat{z}^{d,R}}{2}\right)\,.
\ea
Similar expressions can be written for the leptons.
\begin{figure}[t]
\includegraphics[width=5cm, angle=-90]{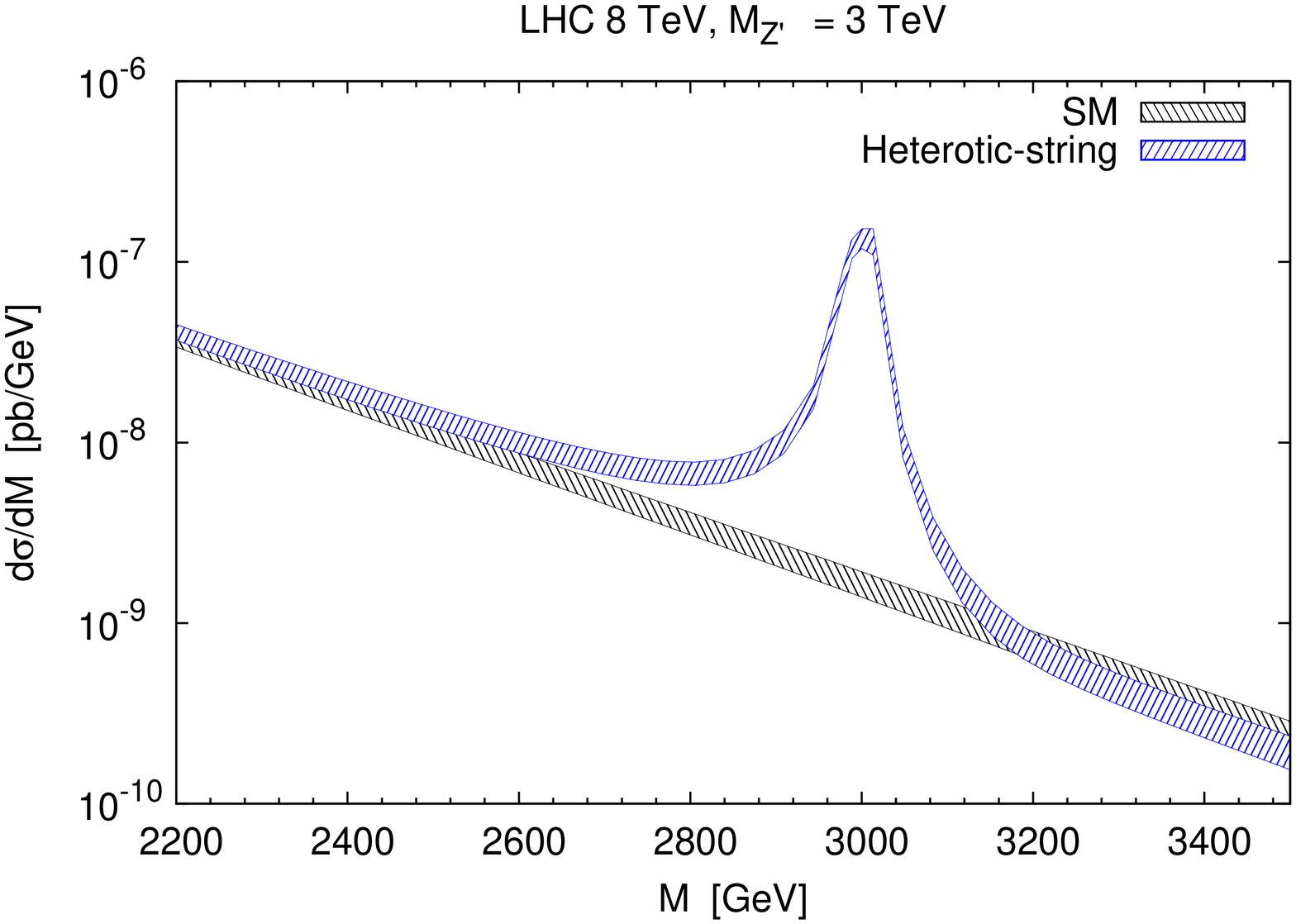}
\includegraphics[width=5cm, angle=-90]{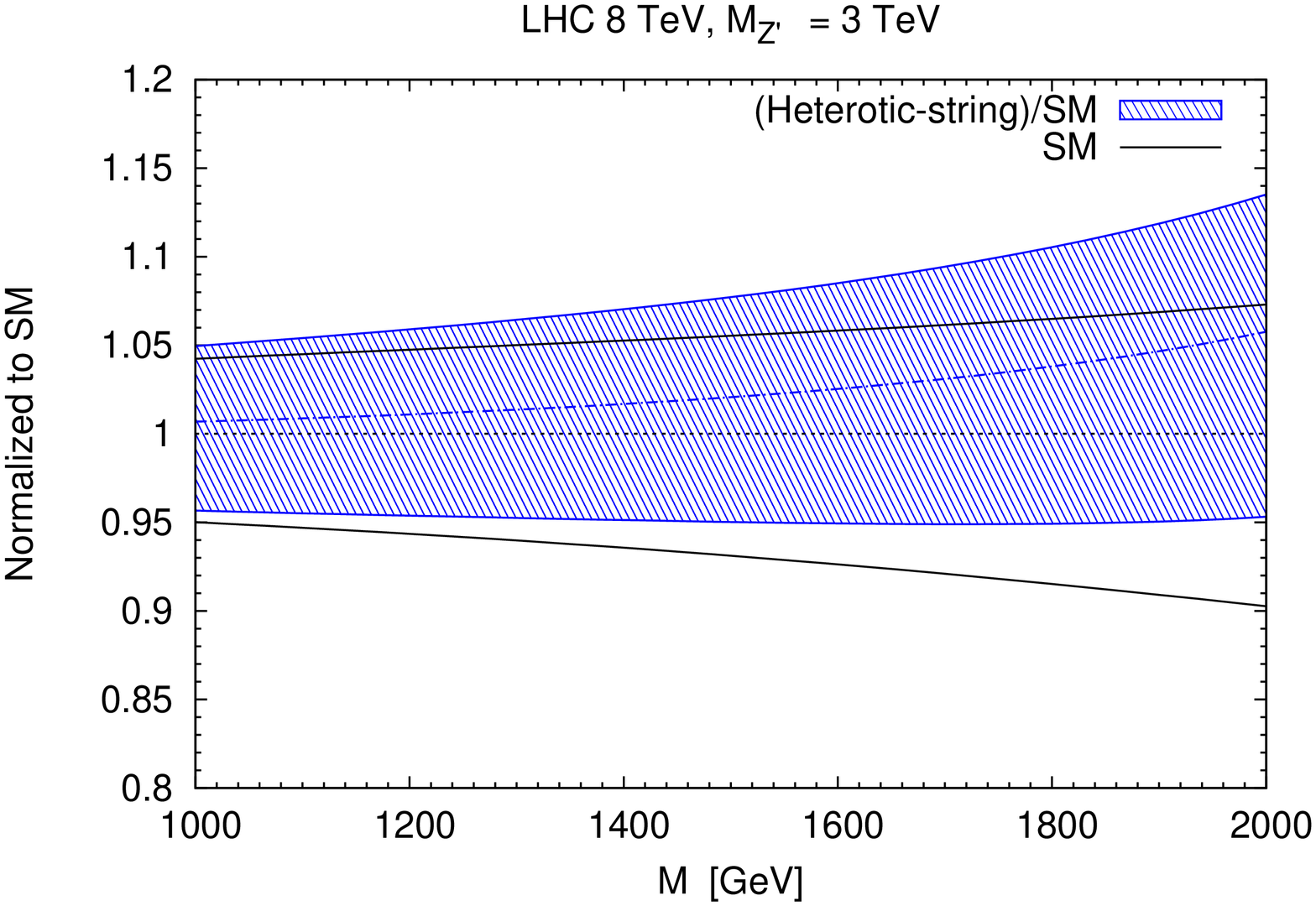}
\caption{Left: LHC 8 TeV DY invariant mass distribution at NNLO for a case ii) 
high $B-L$ breaking scale $Z'$. 
$M_{Z^{\prime}}=3$ TeV, $\tan\beta=10$, and $Z'$ coupling $g_z=g_Y$. 
Hatched bands represent PDFs + scale uncertainties added in quadrature. 
Right: same as in Left, but normalized to the SM.}
\label{fig1}
\end{figure}
 
The main phenomenological results for the high $B-L$ breaking scale model are illustrated in Fig.~\ref{fig1}. 
The left figure shows the $Z'$ invariant mass distribution (blue) compared to the SM (black) 
while, in the right figure, the same prediction is normalized to that of the SM in the 1 to 2 TeV mass range.
Bands with different hatching represent the sum in quadrature of the 
uncertainty relative to the CT14NNLO PDFs~\cite{Dulat:2015mca} rescaled to the 68\% C.L., 
plus the uncertainty associated to independent variations of the $\mu_F$ and $\mu_R$ scales. 
Different choices for the PDFs, obtained from recent analyses~\cite{nnpdf3.0,MMHT,ABM} 
including LHC run-I measurements, give similar results. 
The heterotic-string prediction is almost indistinguishable from the SM in the 1 TeV mass region, 
and deviations start to be more evident around 2 TeV where the central value starts to rise.
In the high-mass region far from the resonance, the SM central value is larger than the heterotic-string prediction, 
but there is a substantial overlap between the two uncertainty bands.
The decay width of the $Z'$ predicted by this model is $\Gamma_{Z'} = 8.76$ GeV that 
is more than three times larger than that of the SM $Z_0$.

\section{Conclusions}

In this paper we surveyed the possibility of low scale $Z^\prime$s and 
$W^\prime$s in three generation heterotic--string vacua. The 
semi--realistic free fermionic models produce the Standard Model spectrum
and the necessary Higgs states for viable symmetry breaking and 
fermion mass generation. The models possess the $SO(10)$ embedding of the 
Standard Model state and the $SO(10)$ normalisation of the weak--hypercharge. 
Hence, they can reproduce viable values of $\sin^2\theta_W(M_Z)$ and 
$\alpha_s(M_Z)$. These are the first order criteria that a viable string vacuum
should possess. The next major constraints on the string models 
are proton stability and suppression of left--handed neutrino masses. 
These two constraints are in tension because on the 
one hand proton stability prefers a low scale $U(1)_{Z^\prime}$ 
breaking, whereas suppression of neutrino masses works more naturally 
with a high scale  $U(1)_{Z^\prime}$ breaking.
As we discussed in section \ref{case1} low scale $U(1)_{Z^\prime}$ breaking
requires the introduction of the ad hoc VEV $\langle\phi\rangle\sim 1$keV. 
The alternative ${\cal Z}^\prime$ discussed in section \ref{case2}
uses a high scale $U(1)_{Z^\prime}$ breaking but anomaly cancellation
necessitates the augmentation of the spectrum into complete $27$ 
multiplets, potentially generating new proton decay operators. 
We further remark that while field theory models allow much more 
model building 
freedom, the straitjacket imposed by synthesising the Standard Model
with gravity in the framework of string theory is by far more restrictive. 

Each of the cases discussed in section \ref{lightu1s} has 
a distinct signature. Case I in section \ref{case1} has 
an additional $Z^\prime$ but no additional states charged
under the Standard Model, with the only additional particles
being the three right--handed neutrinos. Case II of section
\ref{case2} requires the existence of additional colour 
triplets and electroweak doublets in the vicinity of the 
${\cal Z}^\prime$ breaking scale. Case III of the 
left--right symmetric models of section \ref{case3}
contains $W^\prime$s in addition to $Z^\prime$. 
Similarly, case IV of section \ref{case4} 
gives rise to additional vector 
bosons from the $SU(4)$, and $SU(2)_R$ group factors. 
Case V in section \ref{case5} 
produces the $SU(4)$ vector bosons but not 
the $SU(2)_R$. Finally, in the models of case IV with a leptophobic
$Z^\prime$ or custodial $SU(2)$ the additional vector
bosons couple to either the quarks or the leptons 
but not to both. 

For case II of section~\ref{case2} we studied 
the NNLO Drell-Yan invariant mass distribution at the LHC 8 TeV for 
a $Z'$ with mass $M_{Z'}$ = 3 TeV, and estimated the 
main sources of uncertainty in the QCD theory prediction.  
The uncertainty associated to the partonic content of the proton 
is the dominant one and is a limiting factor for precision at the present time.

Observations of one or more additional vector bosons
at the LHC will choose the right model or eliminate all of the 
above, and in fact, the majority of semi--realistic string models 
constructed to date.
Furthermore, the observation of additional vector bosons at the LHC will
restrict the exploration of string vacua, and will elevate 
the utility of high-energy dilepton pair production at hadron colliders.

\section*{Acknowledgments}

AEF thanks theoretical physics groups at CERN and Oxford University 
for hospitality.
This work was supported in part by the STFC (ST/L000431/1) and 
by the Lancaster-Manchester-Sheffield Consortium for Fundamental Physics under STFC grant ST/L000520/1.

\end{document}